\documentclass[a4,useAMS,usenatbib,usegraphicx]{mn2e}
\usepackage{subfig}
\usepackage{lscape}
\usepackage{natbib}
\usepackage{amssymb,amsmath}
\usepackage{fixltx2e}
\usepackage{longtable}
\usepackage[T1]{fontenc}
\usepackage{hyperref}
\hypersetup{
    colorlinks=true,
    linkcolor=blue,
    filecolor=magenta,      
    urlcolor=blue,
}

\title[Turbulence in HI clouds]{Simulating the evolution of optically dark HI clouds in the Virgo cluster : will no-one rid me of this turbulent sphere ?}
\author[R. Taylor, R. W\"{u}nsch, J. Palou\v{s}]{R. Taylor$^{1}$\thanks{Email: rhyst@naic.edu},  R. W\"{u}nsch$^1$, J. Palou\v{s}$^1$\\
$^1$Astronomical Institute of the Czech Academy of Sciences, Bocni II 1401/1a, 141 31 Praha 4\\}

\begin{document}

\newcommand{\HI}{H\textsc{i}}
\newcommand{\Msolar}{M$_{\odot}$}
\newcommand{\kms}{km\,s$^{-1}$}

\date{2016}

\pagerange{\pageref{firstpage}--\pageref{lastpage}} \pubyear{2016}

\maketitle

\label{firstpage}

\begin{abstract}
Most detected neutral atomic hydrogen (\HI{}) at low redshift is associated with optically bright galaxies. However, a handful of \HI{} clouds are known which appear to be optically dark and have no nearby potential progenitor galaxies, making tidal debris an unlikely explanation. In particular, 6 clouds identified by the Arecibo Galaxy Environment Survey are interesting due to the combination of their small size, isolation, and especially their broad line widths atypical of other such clouds. A recent suggestion is that these clouds exist in pressure equilibrium with the intracluster medium, with the line width arising from turbulent internal motions. Here we explore that possibility by using the FLASH code to perform a series of 3D hydro simulations. Our clouds are modelled using spherical Gaussian density profiles, embedded in a hot, low-density gas representing the intracluster medium. The simulations account for heating and cooling of the gas, and we vary the structure and strength of their internal motions. We create synthetic \HI{} spectra, and find that none of our simulations reproduce the observed cloud parameters for longer than $\sim$100 Myr : the clouds either collapse, disperse, or experience rapid heating which would cause ionisation and render them undetectable to \HI{} surveys. While the turbulent motions required to explain the high line widths generate structures which appear to be inherently unstable, making this an unlikely explanation for the observed clouds, these simulations demonstrate the importance of including the intracluster medium in any model seeking to explain the existence of these objects.

\end{abstract}


\section{Introduction}
\label{sec:intro}
Although not all nearby galaxies contain significant amounts of neutral atomic hydrogen (\HI{}), it appears that virtually all detected \HI{} at low redshifts is associated with galaxies. The first all (southern) sky \HI{} survey, the \HI{} Parkes All Sky Survey (HIPASS), found no convincing isolated optically dark \HI{} clouds (that is, \HI{} structures clearly detached from any galaxies and with no visible optical counterpart) in its 4,315 detections (\citealt{doyle}), suggesting that gas in dark matter halos is efficiently converted into stars. However the lack of optically dark but gas-rich galaxies might plausibly be attributed to the effects of sensitivity and confusion, since HIPASS suffered from a relatively large 14$'$ beam and poor 13 mJy \textit{rms} sensitivity (\citealt{barnes}). More recent surveys, such as ALFALFA (Arecibo Legacy Fast Arecibo L-band Feed Array; \citealt{alfalfa}) and AGES (Arecibo Galaxy Environment Survey; \citealt{auld}), have made order-of-magnitude improvements over HIPASS yet virtually all (extragalactic) \HI{} clouds still appear to be associated with optically bright galaxies.

The Arecibo-based surveys benefit from the larger size of the dish, with a beam size of 3.5$'$ that greatly reduces confusion by reducing the number of potential optical counterparts. They also achieve much greater sensitivity, with ALFALFA typically reaching 2.2 mJy and AGES 0.6 mJy \textit{rms}. While not covering as large an area as HIPASS, they still cover the full gamut of galaxy environments, from voids up to rich galaxy clusters. \cite{cannon} report that $<$1.5\% of the ALFALFA detections could not be easily matched with an optical counterpart, and only around 0.3\% of their detections cannot easily be explained as gas which has simply been removed from a parent galaxy. Similarly, using AGES \cite{me12} and \cite{me13} reported a total of eight such clouds in a 20 square degree region of the Virgo cluster. With (at the time) only two additional other dark detections in the whole of AGES, the fraction of dark clouds is no more than about 1\%  (but see below).

Despite this, the nature and origin of \textit{some} of the clouds in the Virgo cluster is very difficult to explain. Tidal debris has generally been the most popular explanation, since the models of \cite{bekki} and \cite{duc} showed that large amounts of \HI{} could be removed to large distances from its parent galaxy by tidal encounters. We confirmed in \cite{me16} and \cite{me17} (hereafter T16 and T17 respectively) that while this explanation is perfectly reasonable for most clouds (see T16 for a review), others remain puzzling. Specifically, we were able to explain features of relatively low velocity width ($<$\,50\,\kms{}, which is typical of most of the known clouds) using pure gravitational harassment (multiple high-velocity tidal encounters, \citealt{mooreharas}). The difficulty arises in explaining the more unusual clouds in the Virgo cluster discovered by AGES, which are isolated ($>$100 kpc from the nearest galaxy), small ($<$17 kpc diameter), and most unusually of all, have very high velocity widths ($\sim$150 \kms{}). They also show no signs of extended \HI{} emission. T16 and T17 found that although these features could be reproduced, they were extremely rare in the simulations and would never be detectable for more than a few tens of megayears. In contrast, T16 also showed that dark matter-dominated, gas rich, optically faint or dark galaxies with parameters matching the observed clouds could be a more viable explanation, as in simulations they are stable over timescales of gigayears.

Recently, new simulations have revived the idea of dark matter dominated halos as a solution to the long-standing `missing satellite' problem, in which cosmological simulations predicted around ten times as many galaxies as are actually observed (e.g. \citealt{moore}). \cite{sawala} describe simulations in which many of the dark matter subhalos contain too few baryons to be detectable. \cite{brooks} reach a similar conclusion and also note that the baryons may not probe the full rotation curve of the dark matter halo - perhaps also explaining the difference in predicted and observed velocity functions (\citealt{klypin}). The theoretical possibility of - even need for - the existence of optically `dark galaxies' therefore remains strong.

On the observational front, there has been a resurgence of interest in low surface brightness (LSB) galaxies, with large populations of physically large but extremely optically faint objects being discovered in the Fornax, Coma and other clusters (e.g. \citealt{vandok}, \citealt{koda}, \citealt{munoz}, \citealt{van16}.). While \cite{vandok} and \cite{koda} speculated that these galaxies require high dark matter fractions to survive in the cluster environment, \cite{bee} measured the kinematics of one such object and found it to be in broad agreement with these predictions and that it has an unusually high dark matter to baryonic mass ratio (see also \citealt{d44}) It seems a natural extension that if such dark matter-dominated low surface brightness galaxies exist, there could also be a population of totally optically dark `no surface brightness' objects in the Universe. 

If the \HI{} clouds in Virgo are indeed generally dark matter-dominated galaxies similar to that measured by \cite{bee} and/or \cite{d44}, they could alter the surprisingly low scatter hitherto observed in the Tully-Fisher relation. This tight relation has long been a puzzle : galaxies of low surface brightness should not follow the same TFR as brighter galaxies unless there is a compensating, unexpected change in mass to light ratio (\citealt{zwaan}). The \cite{bee} result is evidence that this compensating change does not always occur. The \HI{} clouds described in \cite{me12} and \cite{me13} are similar in the sense that their velocity width (and so implied dark matter content) is higher than expected given their baryonic masses (furthermore we cautiously note that the galaxies described in \citealt{aaudgs} may show the \textit{opposite} trend, having uncharacteristically low velocity widths compared to other objects of similar baryonic mass\footnote{Though there are various caveats : the inclination angles are somewhat uncertain and there may be selection effects against detecting high velocity width objects - a detailed discussion is beyond the scope of this paper.}).

In addition, the detection rate of the optically dark \HI{} clouds is low overall but a strong function of environment. In particular, 22 clouds were detected around M33, which is sufficient to explain the lack of its optically bright satellite galaxies (\citealt{olivia}), while at least 62 other clouds have been detected within the Local Group that share similar characteristics expected for the missing satellites (\citealt{adams}). Hence the origin of the clouds as either tidal debris or primordial dark galaxies remains an open question, despite their overall detection rate being far lower than the latter model predicts (\citealt{d04}).

Recently, \cite{BL} (hereafter BL16) have proposed a third option : the clouds in the cluster could be pressure confined by the intracluster medium (ICM). In this scenario the clouds might be stripped from their parent galaxies by ram-pressure stripping, harassment, or a combination of both, but instead of collapsing or dispersing they then reach pressure equilibrium with the ICM. However they need not be stable indefinitely. BL16 note that their density would be close to that needed for star formation, which could also explain why many similar clouds turn out to have faint optical counterparts (\citealt{cannon}).
 
Although such pressure-confined clouds need not be stable over, say, a Hubble time, the stability timescale is still an important issue for any proposed explanation of the \HI{} clouds - if too short then there would be very little chance of detecting them. This is especially important given the number of detections (eight in total) and lack of associated streams which we might expect to be present if they are indeed non-primordial objects. The latter point is discussed in detail in T17.

Previous numerical modelling of such clouds is rare. Intriguingly, \cite{vilnar} found that a few star-free \HI{} clouds were produced in their simulations of cluster formation. However as clouds of the observed masses would only have around 10 particles in their simulations, it is unclear whether they are simply a numerical artifact or what their measured velocity widths would be. The latter point is crucial, as we showed in T16 and T17 that while low-width clouds ($<$50 \kms{}) of small size ($<$ 20 kpc diameter) are readily produced by harassment (almost ubiquitous in our simulations), high-width~clouds ($>$100 \kms{}) of that size are extremely rare. Thus for a model which can explain the low-width clouds, it is not necessarily a matter of merely minor adjustments to explain the high-width clouds~: such a model could potentially be correct for the low-width clouds but simply wrong for clouds of high velocity width. Hence here we concentrate on the high-width clouds as, while they are not at all typical compared to other \HI{} clouds, they appear to be much more difficult to explain with the conventional `tidal debris' scenario. Furthermore, while \cite{bell} have recently convincingly demonstrated that low width clouds could remain in stable pressure equilibrium with the ICM, the high width clouds require very different conditions, which we shall discuss more in section \ref{sec:turbulenceisawesome}.

In this paper we use the grid-based hydrocode FLASH (\citealt{fryxell}) to analyse the behaviour of such pressure-supported clouds at high resolution. Our goal is to quantitatively estimate the duration for which such clouds would remain similar to the AGES detections, especially with regards to their small size and high line width. We vary the size, mass, velocity dispersion and metallicity of the clouds, and account for heating and cooling by different effects from the intracluster medium. We do not consider the possible formation mechanism of such clouds, which would require a more complicated setup.

The rest of this paper is organised as follows. In section \ref{sec:sims} we describe the exact simulation setup and the methods used to analyse the results. We assess the simulations in section \ref{sec:results} and discuss our conclusions in section \ref{sec:conc}.

\section{Simulation Setup}
\label{sec:sims}

The model proposed in BL16 is that the clouds are in stable pressure equilibrium with the ICM, based on their observed line widths and density of the ICM at their locations\,: expansion pressure from the cloud is balanced by compressional pressure from the ICM. This expansion pressure must be dynamic, i.e. turbulence, as the thermal pressure needed for equilibrium would require temperatures well in excess of 100,000\,K - much too hot for the gas to remain as \HI{}. Turbulent motions make the evolution of the properties of the cloud (size, structure, temperature) difficult to predict analytically. We do not expect a global equilibrium state to develop, since, unlike in the case of thermal pressure, different parts of the cloud must have different pressures due to their different velocities. Instead, we here numerically investigate how long the clouds would have properties in agreement with the AGES observations before either dispersing or collapsing to form stars. We attempt to determine the initial conditions that could allow the clouds to match the observations (see sections \ref{sec:matchobs} and \ref{sec:matchobs2}) of both the S/N and line width for the longest possible time.
  
Observationally, we are limited to just two direct measurements of the clouds themselves : the velocity width from their \HI{} line profiles (typically 150 \kms{}) and their \HI{} masses (around 2$\times$10$^{7}$\Msolar). We also have the constraint that since the objects are not spatially resolved by Arecibo, they cannot be much larger than the Arecibo beam - around 17 kpc diameter at the 17 Mpc assumed distance of the Virgo cluster. Since the BL16 model is that the clouds are purely gaseous, we neglect any stellar component - this also appears to be observationally justified, as discussed in T16. The lack of optical emission allows us to set a probable lower size limit on the objects : if they were smaller than about 2 kpc diameter, they would have column densities comparable to star-forming dwarf galaxies so we would expect them to have easily detectable optical counterparts. Full observational parameters for the clouds are given in T16 table 3. Below, we discuss how we relate these observational constraints and measurements to our simulations.  
   
Our models use a 3D grid containing a spherical Gaussian density distribution for the warm \HI{} gas surrounded by a uniform density hot ICM. In all cases, the simulation domain is a cube 32 kpc on a side (about twice the Arecibo beam width - sufficient to determine if the emission would become resolved) with 512$^{3}$ cells, with each cell being 62.5 pc across. This resolution allows us to predict what we might detect with higher resolution observations if the clouds are indeed pressure-supported spheres (they are unresolved point sources in the Arecibo data). All simulations are allowed to run for 400 Myr. The calculations were carried out on cluster Salomon of the Czech National Supercomputing Centre IT4I \footnote{http://www.it4i.cz/?lang=en}; the computational time was obtained under project OPEN-9-25.

\subsection{Numerical code}
\subsubsection{Hydro solver and gravity}
Our numerical model is based on the three-dimensional, adaptive mesh refinement (AMR) code \textsc{flash} v4.3 (\citealt{fryxell}). The AMR is handled by the PARAMESH library (\citealt{paramesh}). The whole code is parallelised via domain decomposition under the Message Passing Interface (MPI). The hydrodynamic equations are solved using a modified version of the Piecewise Parabolic Method of \cite{colella} with the time-step controlled by the Courant-Friedrichs-Lewy criterion. Self-gravity of the gas is calculated using the tree code algorithm described in \cite{wunsch}. We neglect the tidal field of the cluster and its member galaxies, i.e. the external gravitational field. Ultimately incorporating this effect, together with the ram pressure of the ICM may be necessary for understanding the formation and evolution of the clouds, but this is technically challenging and well beyond the scope of the current work.

\subsubsection{Sink particles}
We use the \textsc{flash} sink particles module described in \cite{fed}, which enables treatment of gravitationally unstable collapsing gas. If the gas density in a certain grid cell exceeds a threshold density $\rho_\mathrm{sink}$, and if the gas within the so-called accretion radius (corresponding to $2.5\times$ the grid cell size), fulfills a number of criteria, then a sink particle is created. These criteria are : (i) the cell is at the highest refinement level; (ii) the cell represents a local minimum of the gravitational potential; (iii) the mass below the accretion radius exceeds the Jeans mass; (iv) the flow is converging; (v) the gas is gravitationally bound; (vi) the region does not overlap with some other sink particle. Additionally, a fraction of gas with density exceeding $\rho_\mathrm{sink}$ within the accretion radius of each particle is accreted onto it, i.e. the gas density is truncated to $\rho_\mathrm{sink}$ and the mass is added to the mass of the sink particle. We set $\rho_\mathrm{sink} = 1.6\times 10^{-25}$\,g\,cm$^{-3} \sim (0.2 \mathrm{km/s}\times 62.5\,\mathrm{pc})^2 / G$ for all models in this work, and even though this density threshold is occasionally exceeded, no sink particles were formed in any of the presented models, i.e. at least one of the other criteria was not fulfilled. 

\subsubsection{Cooling and heating}
The gas in the simulations is subject to radiative cooling and heating. Cooling of the gas with temperature above $10^4$\,K is calculated using the table by \citet{schure} and the chemical composition is either solar \cite{anders} or 1\% of solar (all elements heavier than helium have abundances decreased by factor $100$). For gas at temperatures below $10^4$\,K, cooling is determined using the table by \citet{dgmc} which requires the ionisation degree of the gas. Ideally this could be calculated using a chemical network, but for simplicity we instead use a power-law fit to results obtained by \citet{stern} using the methods described in \cite{wolf}. These estimates were obtained  for \HI{} in dark matter mini-halos (and high velocity clouds) embedded in the hot gas of the Milky Way halo and photoionised and heated by the background metagalactic field, assuming that the physics is similar. The logarithm of the ionisation degree, $x_e$, is given by the following function of the logarithm of the hydrogen nuclei number density $\xi = \log(n_\mathrm{H})$
\begin{equation}
\log(x_e) =
\left\{
\begin{array}{lcl}
-1 & \mathrm{for} & \xi < -3 \\
-1 - (\xi+3)/2 & \mathrm{for} & -3 < \xi < -1.5 \\
-1.75 - (\xi+1.5)/8 & \mathrm{for} & -1.5 < \xi < 0.5 \\
-4 & \mathrm{for} & \xi > 0.5 \\
\end{array}
\right. 
.
\end{equation}
Heating is assumed to be a constant, uniform value of $10^{-28}$\,erg\,s$^{-1}$\,H$^{-1}$. This value comes is an approximation to the results of the analysis by \cite{stern} - see their Fig. 4. It combines heating by the UV and X-ray radiation from the surrounding hot gas, photoelectric heating, and heating by H$_2$ photodissociation. Cooling and heating terms are included into the energy equation by the integration procedure using sub-cycling with sub-steps determined to follow the temperature values recorded in the cooling table, similar to the one described in \cite{zhu}. We do not consider the possibility of heat conduction from the ICM due to the technical difficulties posed in doing so. 

\subsection{Initial conditions}
\subsubsection{Cloud density and temperature}
Given the observational constraints we use two sizes for the clouds : a radius of 2 or 8 kpc, taking 8 kpc for our fiducial model. In most cases we use a mass of 2.1$\times$10$^{7}$\Msolar{}, which well approximates the observed mass of the clouds. 

We do not have any observational constraints on the density profile of the clouds so we choose to model them with Gaussian distributions, with the density in the centre being three times higher than at the surface. This is fully compatible with the BL16 model of the clouds and, as we shall see, the density of the clouds evolves rapidly and strongly due to their high velocity dispersion. It is therefore very unlikely that changing the initial density structure of the clouds would have a significant effect on the final results.

We do not have any observational temperature constraint on the clouds so we assume 10$^{4}$\,K for the initial temperature. This agrees with the typical $\sim$10\,\kms{} velocity dispersion of \HI{} gas in galaxies. 

\subsubsection{Initial (turbulent) velocity field}
\label{sec:turbulenceisawesome}
As discussed in BL16 there are three main sources of pressure that could support the clouds against collapse and so prevent star formation : thermal pressure, rotation, and turbulence. Since the BL16 model neglects rotation we do not include any rotational component of the clouds in these simulations; if the clouds are rotation-dominated this implies a significant dark matter component, which we discuss at length in T16. The thermal pressure for our assumed initial temperature corresponds to a line width of 10 \kms{}, far below the observed values of $\sim$150\,\kms{}. This means that the dominant source of the supporting pressure must be dynamic motions, i.e. turbulence.

The recent study of \cite{bell} provides a nicely complementary approach to our own. The authors investigate the survival of clouds with much lower velocity widths ($\sim$30\,\kms{}), which are smaller (1-2 kpc) and also forming stars. Their objects are therefore very different to ours : in particular, they are internally supported only by thermal pressure, which is not possible for the high width objects we examine. Thus the gigayear survival of the \cite{bell} clouds does not imply a similar longevity for the AGES clouds.

Turbulence in the \HI{} is here modelled using the method described in \cite{clark}. The velocity field of the gas in the cloud is a composition of modes with wavenumbers $k$ between $k_\mathrm{min}\times (2\pi/L)$ and $k_\mathrm{max}\times (2\pi/L)$, where $L$ is the size of the computational domain. The mode amplitudes are generated randomly, with a Gaussian distribution and dispersion chosen so that the kinetic energy spectrum is given by the power law $E(k)\!\propto\!k^{\alpha}$. For the fiducial run, the three parameters describing the initial turbulent velocity field have values $\alpha = -5/3$ (corresponding to Kolmogorov turbulence, following \citealt{dcb}), $k_\mathrm{min} = 2$ and $k_\mathrm{max} = 32$. However, we have performed simulations varying all of these parameters from our chosen fiducial conditions.

After the modes are generated, the resulting velocity field is rescaled in order to set the desired initial FWHM, which we take to be 200 \kms{}. This is deliberately slightly higher than the observational value of $\sim$150 \kms{}, as our preliminary tests indicated that the FWHM tends to drop extremely rapidly. This higher value is still essentially compatible with the observations but is designed to give the proposed scenario the best possible chance of success, i.e. to reproduce the observations for the longest possible time. At this velocity width the calculated radius of the clouds for pressure equilibrium would be 8.1 kpc using equations 1-3 from BL16, only slightly higher than our maximum actual value of 8.0 kpc. In practise, we find that the structures generated by turbulence evolve rapidly and no special equilibrium state develops. Hence this small size difference is not expected to make any difference to the end result (additionally, as we shall see, altering the velocity field whilst keeping the FWHM constant also causes significant changes).

No energy is injected to maintain the turbulence as there is no obvious source to maintain it in these pure gas clouds which are $>$\,100\,kpc from the nearest galaxy. We do not attempt to simulate the origin of the turbulence in this work; presumably it might arise during the galaxy-galaxy encounters which are hypothesised to have formed the clouds, or perhaps due to the continuous mixing of a stripped gas wake with the ICM. In any case, injecting additional energy into unbound systems can only cause them to disperse more quickly, though this may be investigated in more detail in a future work.

\subsubsection{Parameters of the ICM}
The turbulent cloud is embedded in a hot rarefied gas representing the ICM. This ambient gas has uniform density and temperature and the values were set, as in T16, according to the prescription of \cite{vol01} which is based on the ROSAT X-ray data described by \cite{schindler}. \cite{vol01} gives a density profile of :
\begin{equation}
\rho = \rho_{C}\Big(1 + \frac{r^{2}}{r_{c}^{2}}\Big)^{-\frac{3}{2}\beta}
\label{eqt:icmrho}
\end{equation}
Where, using values from \cite{vol01}, $\rho$ is the ICM density at a given distance $r$ from the cluster centre, $\rho_{C}$ is the central density (4$\times$10$^{-2}$cm$^{-3}$), $r_{c}$ is the radius of the cluster core (13.4 kpc), and $\beta$ is the slope parameter (0.5). We also use \cite{vol01}'s approximation that the ICM temperature $T$ is a uniform 10$^{7}$ K (see also \citealt{shib}). We assume a clustercentric distance $r$ of 1.5 Mpc, giving a density of 3.4$\times$10$^{-5}$\,cm$^{-3}$ (the real clouds have distances ranging from 1.1 to 1.8 Mpc, or a range of densities from 2.6$\times$10$^{-5}$\,cm$^{-3}$ to 5.5$\times$10$^{-3}$\,cm$^{-2}$).

\subsubsection{Computational domain}
In all cases, the simulation domain is a cube with side $L = 32$\,kpc (about twice the Arecibo beam width - sufficient to determine if the emission would become resolved). For majority of runs, the basic grid has 256$^{3}$ cells and one more level of refinement is added in regions where the gas temperature drops below $10^5$\,K, resulting in each cell being $62.5$\,pc across. This resolution allows us to predict what we might detect with higher resolution observations if the clouds are indeed pressure-supported spheres (they are unresolved point sources in the Arecibo data). In the high resolution run, the grid cell is half of the size in the standard run. All simulations are allowed to run for 400 Myr.

\subsection{Compatibility with the original proposed model}
\label{sec:matchobs}
We have used slightly different parameters to the original BL16 proposal for two reasons. Firstly, BL16 consider (in addition to the high velocity width clouds we explore here) the two low-width clouds, whereas we ignore them in this study. This is because we showed in T16 and T17 that clouds of low velocity width can be readily explained by simple tidal encounters between galaxies, with no need to invoke any more complex physics. Hence our fiducial model was explicitly designed to model the high-width clouds as we consider these to be the most interesting and difficult to explain. Secondly, after deciding on our fiducial model, subsequent models were constructed based on the previous results in order to try (as mentioned) and give this proposed scenario the best chance of matching the observations for the longest possible time.

Despite this, the differences in the initial conditions between our fiducial model and the original BL16 proposal are minor. Our assumed density and temperature of the ICM gives a pressure only 1.4 times greater than that using in BL16, as is our \HI{} mass compared to the median value of the BL16 clouds. The main difference in the parameters compared to BL16 is our chosen line width, which is rather higher than those in BL16 (though as discussed, at our maximum size this gives an internal pressure of the clouds very close to being in apparent equilibrium with the ICM).

\subsection{Analysis methods}
\label{sec:anmeth}
\subsubsection{Synthetic spectra}
When calculating the spectra of the clouds, we assume that the warm gas (with temperatures between $10^3$\,K and $3\times 10^4$\,K) corresponds to \HI{}. Below the lower temperature limit all the gas is assumed to be neutral but molecular (we do not account for the molecular fraction varying with temperature or density, as we do not have a good analytic approximation for this). Above the upper limit the gas is assumed to be completely ionised.

As mentioned we have only two direct observational measurements of the clouds : their mass (equivalent to S/N) and velocity width; plus a third constraint that they must have r\,$<$\,8.5 kpc to be unresolved to Arecibo. This means we need to create synthetic spectra of the clouds, which allow us to monitor the both parameters over time. For the \HI{} line this is relatively straightforward. The  densities (after correcting for molecular and ionised gas) within each cell in the specified region (of the same size as the Arecibo 17 kpc beam) are integrated to compute the total gas mass within that region. We can express the standard equation for \HI{} mass as follows :
\begin{equation}
M_{HI} = 2.36\times10^{5}D^{2}\:V_{chan}\:\sigma_{rms}\:S/N
\label{eqt:himass}
\end{equation}
Where $M_{HI}$ is the \HI{} mass in solar masses (in a single channel), $D$ is the distance to the source in Mpc, $V_{chan}$ is the channel width in \kms{}, $\sigma_{rms}$ is the noise level in Jy, and S/N is the signal-to-noise ratio. This can be trivially re-arranged to compute the S/N for a survey of any given $\sigma_{rms}$, which we here take to be 0.6 mJy to emulate AGES; likewise we set $V_{chan}$ to 10\,\kms{}. Density, temperature and pressure are measured directly from the simulation.

\begin{figure}
\centering
\includegraphics[width=85mm]{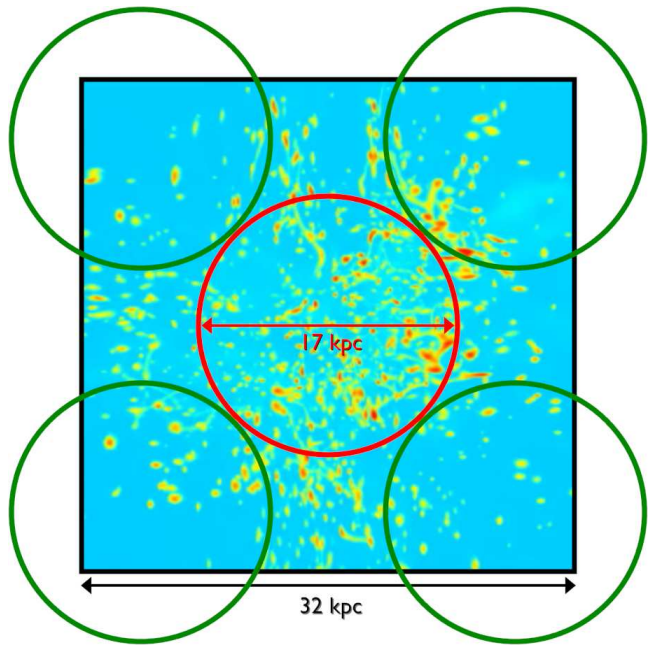}
\caption[hienv]{Illustration of how we compute spectra in the central beam (red) and several independent adjacent side beams (green), with the background showing a column density snapshot of our fiducial run.}
\label{fig:beams}
\end{figure}

Ideally we would convolve the simulated data with a 17 kpc Gaussian function in order to create synthetic pixel maps which we could examine to see if the clouds would be resolved (or detected in multiple pixels) given Arecibo's spatial resolution. This is not practical given the computational expense. Instead we adopt the simpler approach of examining the gas in multiple, independent regions, as shown in figure \ref{fig:beams}. The `side beams' (not to be confused with \textit{sidelobes}) are placed diagonally rather than along the vertical and horizontal axes - this would reduce their area inside the simulation domain to 50\%, whereas the diagonal configuration increases this to 60\%. This is not ideal, however in practise the peak S/N in the side beams rarely even exceeds 1.0, so the possibility of the clouds becoming resolved to Arecibo is essentially a non-issue. 

The line profile evolution depends on the viewing angle. We compute these from three orthogonal directions for the central beam and generally show the average of the three directions, unless otherwise stated. Profiles for the side beams are calculated along a single direction (the z axis).

\subsubsection{Comparison with observations}
\label{sec:matchobs2}
Since we want to measure how long the clouds are a good match to the observations, we must first define what we consider to be a good match given the variation in the observed cloud parameters. The practical problems of detecting \HI{} in real data cubes, especially given the imperfect baselines and various artifacts in the noise, are discussed in detail in \cite{me15} (see also \citealt{me12}, \citealt{duchamp}, \citealt{saintonge}). Briefly, a source found in a single pixel in a single velocity channel at a S/N of 4.0 would almost certainly not be detected, as given the size of real data cubes, such features occur at a high rate even from perfectly Gaussian noise. Human inspection does not notice such spikes; algorithms have to ignore them to avoid detecting thousands of spurious sources. However, a feature at the same S/N level contiguously extended over many different channels or spatial pixels would have a much better chance of being detected, both by human visual inspection and by automated methods. Yet if the same flux were distributed over close but non-contiguous channels and/or pixels, it would appear as a collection of individual point sources and suffer the same problems as before. In short, it is far from trivial to quantify strict detection limits even using S/N, let alone by total \HI{} mass - a degree of subjectivity is unavoidable. Fortunately, both the simulations and the observations allow us to simplify the problem for our goals here. 

Firstly, in practice, the simulations do not show very narrow spectral peaks, i.e. the single channel spikes that are so problematic for source extraction in observational data sets, but instead tend to show much broader profiles. Moreover, we are explicitly only interested in broad line width features here, since this is one of the most interesting aspects of the AGES clouds that has proven difficult to explain through tidal encounters (T16 and T17). Taken together, it is reasonable to assume that any 4$\sigma$ peak with a FWHM $\geq$ 100 \kms{} would be readily detectable. Features narrower than this threshold would anyway be regarded as very different from the observations and not considered to be a good match.  Secondly, we are explicitly only interested in isolated sources with a total diameter smaller than the Arecibo beam. In other words, given the small size of our simulation volume, we can reject any cloud that is more than 17 kpc across or if there is another detection present at any other spatial location in our simulated volume.

As we set the initial detectable mass to be close to the sensitivity limit of the survey, detectable mass and S/N ratio are implicitly constrained by the simulation and do not require additional restrictions. The detectable mass cannot fall very much without becoming completely undetectable, nor can it rise substantially given the slow rate of cooling of the hot gas. While the S/N cannot fall far from its initial value, in principle it could rise much higher - to a maximum value of $\sim$50 if all the \HI{} was in a single 10 \kms{} channel. However, by equation \ref{eqt:himass}, line width and S/N are related (given the mass restriction), therefore we do not impose any limit on the S/N ratio in order for the cloud to be considered a match to the observations. The only way the S/N could become significantly higher than the observations is if the line width dropped well below the 100 \kms{} threshold, hence the line width provides sufficient comparison to the observations by itself.

\section{Results}
\label{sec:results}
We ran a variety of simulations in which we adjust the clouds velocity width, the spatial scale and slope of the turbulence, and the metallicity. The complete properties are given in table 3. We identified three distinct modes in the simulations : 1) Dispersal, in which the clouds expand and become undetectable; 2) Collapse, in which the clouds shrink with their S/N increasing but their line widths falling; 3) Heating, in which the main process affecting detectability is not the dynamics but the change in temperature, with the clouds becoming undetectable as the \HI{} temperature increases and the gas becomes indistinguishable from the ICM. Of course the effects are not entirely mutually exclusive - the centre of a cloud may collapse while the outer parts disperse, for example - so our categories only attempt to identify which effect is dominant. Examples of each of these three overall behaviours are shown in figure \ref{fig:fidev}.

Here we describe the fiducial model in some detail, with the behaviour of the other simulations presented as comparisons. While the evolution of the two major observable parameters (FWHM and S/N ratio) are shown for all models, to examine the evolution of all physical properties of the system (density, temperature, pressure) we invite the reader to consult the supplementary online material showing movies for all the simulations described here.

\begin{table*}
\caption[stab]{Simulation parameters. In all cases the the ICM gas has a density of 3.4$\times$10$^{-5}$\,cm$^{-3}$ and is at a uniform temperature of 10$^{7}$\,K. Heating is assumed to be a uniform constant value of 10$^{-28}$\,erg\,s$^{-1}$. Model A is our fiducial run. Models B-E inclusive are the same as the fiducial but altering a single global parameter of the clouds or the model. Runs F-I vary the parameters of the turbulence. Models J and I are \textit{ad hoc} adjustments to the fiducial model in an attempt to more closely reproduce the observed line profiles of the real \HI{} clouds. To aid readability, all differences from the fiducial parameters are highlighted in bold.}
\label{tab:simparams}
\begin{tabular}{c c c c c c c c c }
\hline
  \multicolumn{1}{c}{Model} &
  \multicolumn{1}{c}{Mass/10$^{7}$\Msolar{}} &
  \multicolumn{1}{c}{Radius/kpc} &
  \multicolumn{1}{c}{Metallicity/Z$_{\odot}$} &
  \multicolumn{1}{c}{FWHM/\kms{}} &
  \multicolumn{1}{c}{$\alpha$} &
  \multicolumn{1}{c}{$k_{min},k_{max}$} &
  \multicolumn{1}{c}{Resolution} &
  \multicolumn{1}{c}{Result} \\
\hline
  A & 2.1 & 8.0 & 1.0 & 200 & -5/3 & 2,32 & 512$^{3}$ & Dispersal\\
\hline  
  B & 2.1 & \textbf{2.0} & 1.0 & 200 & -5/3 & 2,32 & 512$^{3}$ & Dispersal \\
  C & 2.1 & 8.0 & \textbf{0.01} & 200 & -5/3 & 2,32 & 512$^{3}$ & Heating\\
  D & 2.1 & 8.0 & 1.0 & \textbf{380} & -5/3 & 2,32 & 512$^{3}$ & Dispersal\\
  E & 2.1 & 8.0 & 1.0 & 200 & -5/3 & 2,32 & \textbf{1024$^{3}$} & Dispersal\\
\hline 
  F & 2.1 & 8.0 & 1.0 & 200 & -5/3 & \textbf{8,128} & 512$^{3}$ & Collapse\\
  G & 2.1 & 8.0 & 1.0 & 200 & -5/3 & \textbf{2,4} & 512$^{3}$ & Dispersal\\
  H & 2.1 & 8.0 & 1.0 & 200 & \textbf{-8/3} & 2,32 & 512$^{3}$ & Dispersal\\
  I & 2.1 & 8.0 & 1.0 & 200 & \textbf{-2/3} & 2,32 & 512$^{3}$ & Collapse\\
\hline 
  J & \textbf{4.2} & 8.0 & 1.0 & \textbf{380} & -5/3 & 2,32 & 512$^{3}$ & Heating\\
  K & 2.1 & \textbf{2.0} & \textbf{0.01} & 200 & -5/3 & 2,32 & 512$^{3}$ & Dispersal\\
\hline
\end{tabular}
\end{table*}

\subsection{The fiducial model}
\begin{figure*}
\centering
\includegraphics[width=180mm]{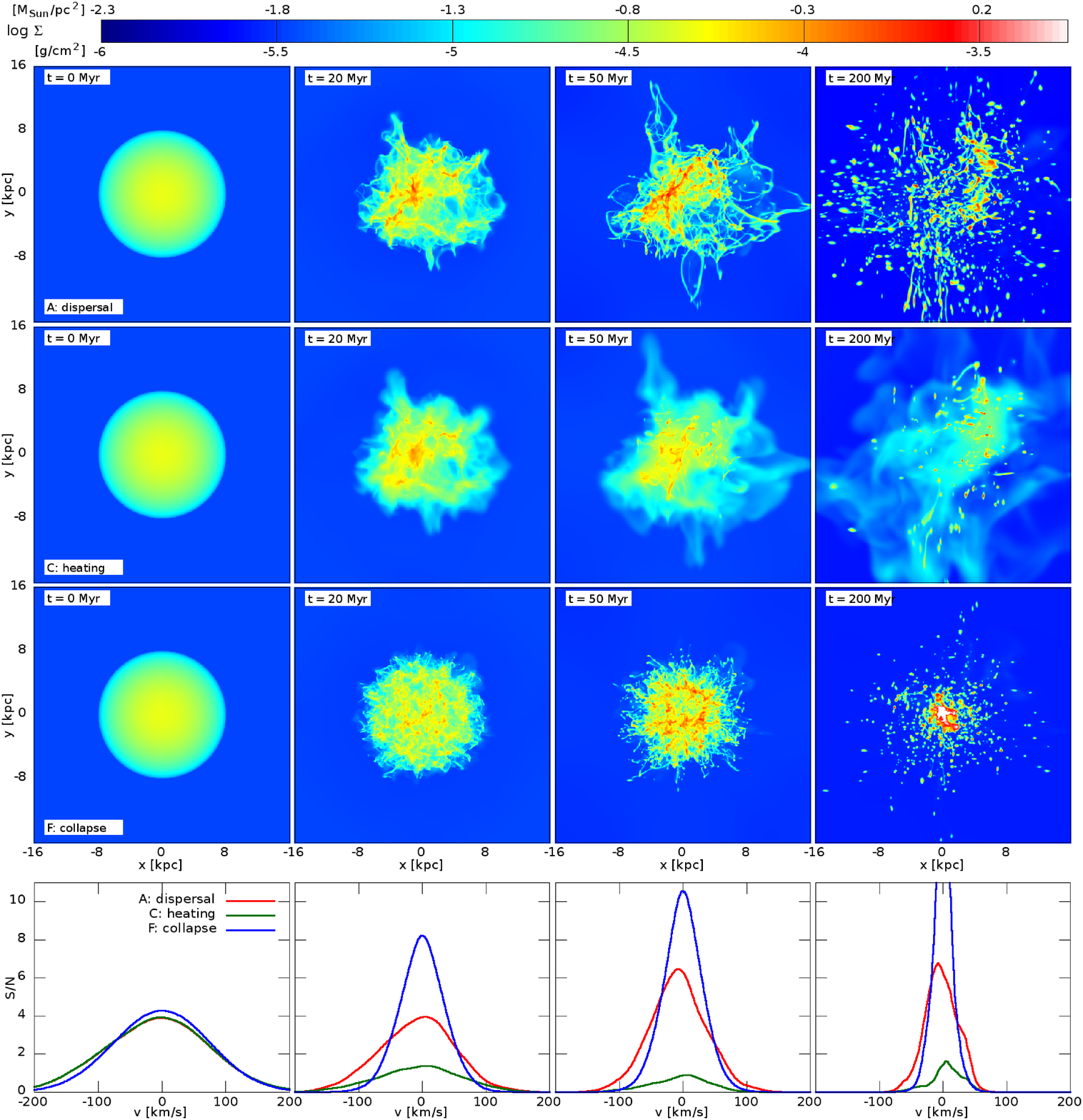}
\caption[hienv]{Snapshots of the evolution of three models demonstrating each of the typical overall behaviours. The upper three rows show the surface densities, where, for reference, the upper value of 10$^{-3.5}$\,g\,cm$^{-2}$ corresponds to 1.9$\times$10$^{20}$\,cm$^{-2}$ or 1.5 \Msolar\,pc$^{-2}$. The line profiles in the bottom row show only the central beam - the side beams never contain detectable gas.}
\label{fig:fidev}
\end{figure*}

Snapshots of the evolution of the column density and line width of the fiducial model are shown in figure \ref{fig:fidev}. The initially smooth Gaussian sphere is rapidly destroyed : within 10 Myr the cloud has transformed into a network of filaments, as various parts of the cloud collide with one another due to their randomised motions. The length of the filaments is comparable to the initial radius of the sphere, though they are much narrower. In this model the cloud has a kinetic to gravitational potential energy ratio $Q$ of almost 900, meaning it is highly unbound. With the relatively large scale of the turbulence, the resulting fragments are free to expand, gradually fragmenting until after approximately 200 Myr the cloud has become a collection of small, discrete dense blobs. By 400 Myr (the end of the simulation) many of the blobs have left the simulation domain, with a remaining mass of warm \HI{} gas of 9$\times$10$^{6}$\,\Msolar.

\begin{figure}
\centering
\includegraphics[width=85mm]{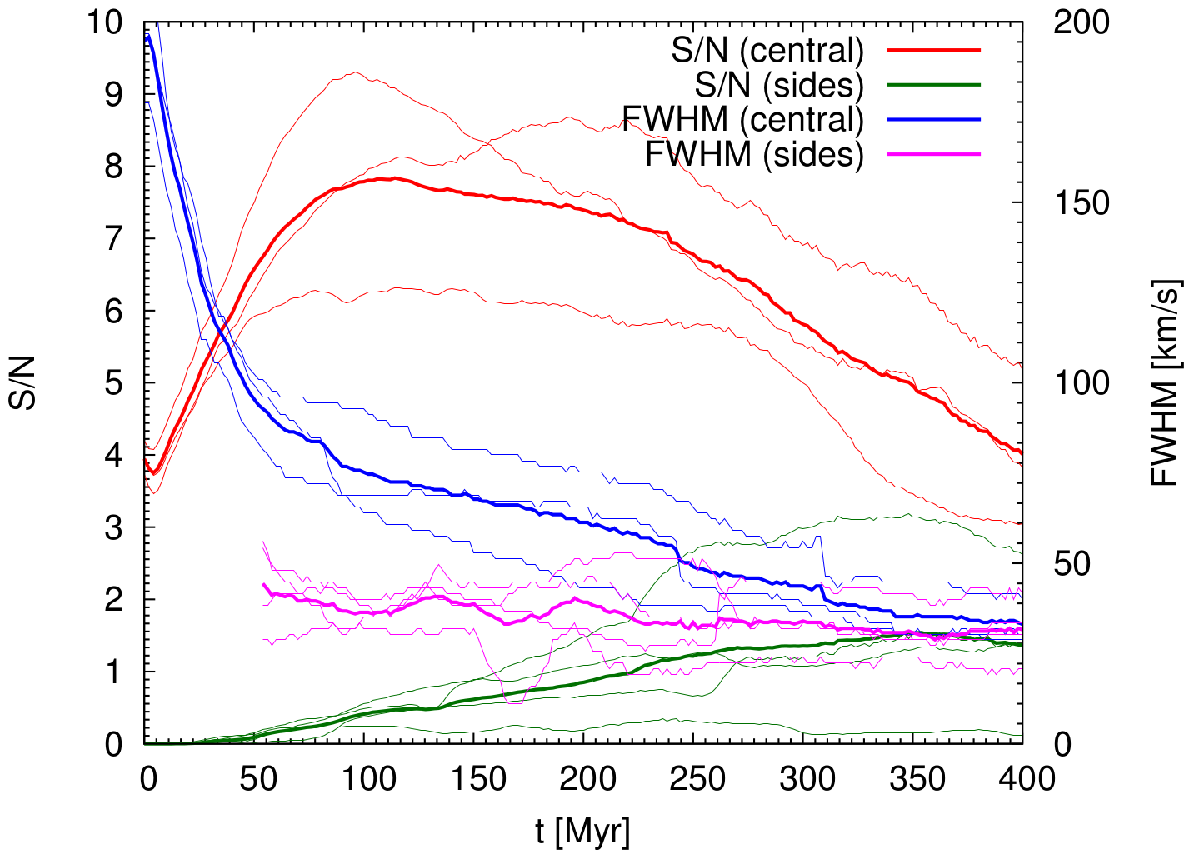}
\caption[hienv]{Snapshots of the line profile evolution of the fiducial model (A). The thick lines indicate the average over all three viewing angles whereas the thin lines show the profiles obtained for separate viewing angles.}
\label{fig:fidprof}
\end{figure}

The line profile shows that the cloud remains detectable even after 400 Myr, though by this point the S/N level appears to be in a terminal decline (see figure \ref{fig:fidprof}). The S/N even increases initially, reaching a peak of about 8 after about 100 Myr and falling back to its initial value of 4 by the end of the simulation. These values are in good agreement with the observed clouds. However, this rise in S/N comes at the expense of the line width. The FWHM drops rapidly, falling below 100 \kms{} in just 50 Myr. After this the FWHM continues to drop approximately linearly, though more slowly, reaching 50 \kms{} after about 240 Myr and 35 \kms{} by the end of the simulation. 

Neither the S/N or the FWHM of the side beams ever approach the observed values of the clouds. The S/N of the side beams did reach 3.0 in one case, but in the others it remained below 1.5. Furthermore the FWHM of the side beams never exceeded 60 \kms{}, well below our target of 100 \kms{}. The evolution of the line profiles gives no hint that the side beams would resemble the observed clouds had we let the simulation continue - the FWHM and S/N are evolving only very slowly at the end of the simulation, and a drop in S/N may be expected as mass continues to leave the simulation volume. Our simulated clouds would always remain unresolved to an Arecibo-sized beam, so in that sense they resemble the observed clouds - but their S/N and FWHM are quite different to the real clouds. 

While there are many caveats regarding the lack of detailed chemistry included in our models, the surface density rarely exceeds 1\,\Msolar\,pc$^{-2}$. This compares favourably with the BL16 prediction that all four of the AGES clouds they consider should be below the 4\,\Msolar\,pc$^{-2}$ threshold at which the \HI{} is expected to transition to H$_{2}$ (though BL16 note that clouds found in other surveys are nearer this threshold), or the approximate 8\,\Msolar\,pc$^{-2}$ believed to be necessary for star formation (e.g. \citealt{cglov}). Unlike the similar mass but greater density cloud described in \cite{bell}, which has some small amount of star formation, these features are expected to remain optically dark.

\begin{figure*}
\centering
\includegraphics[width=180mm]{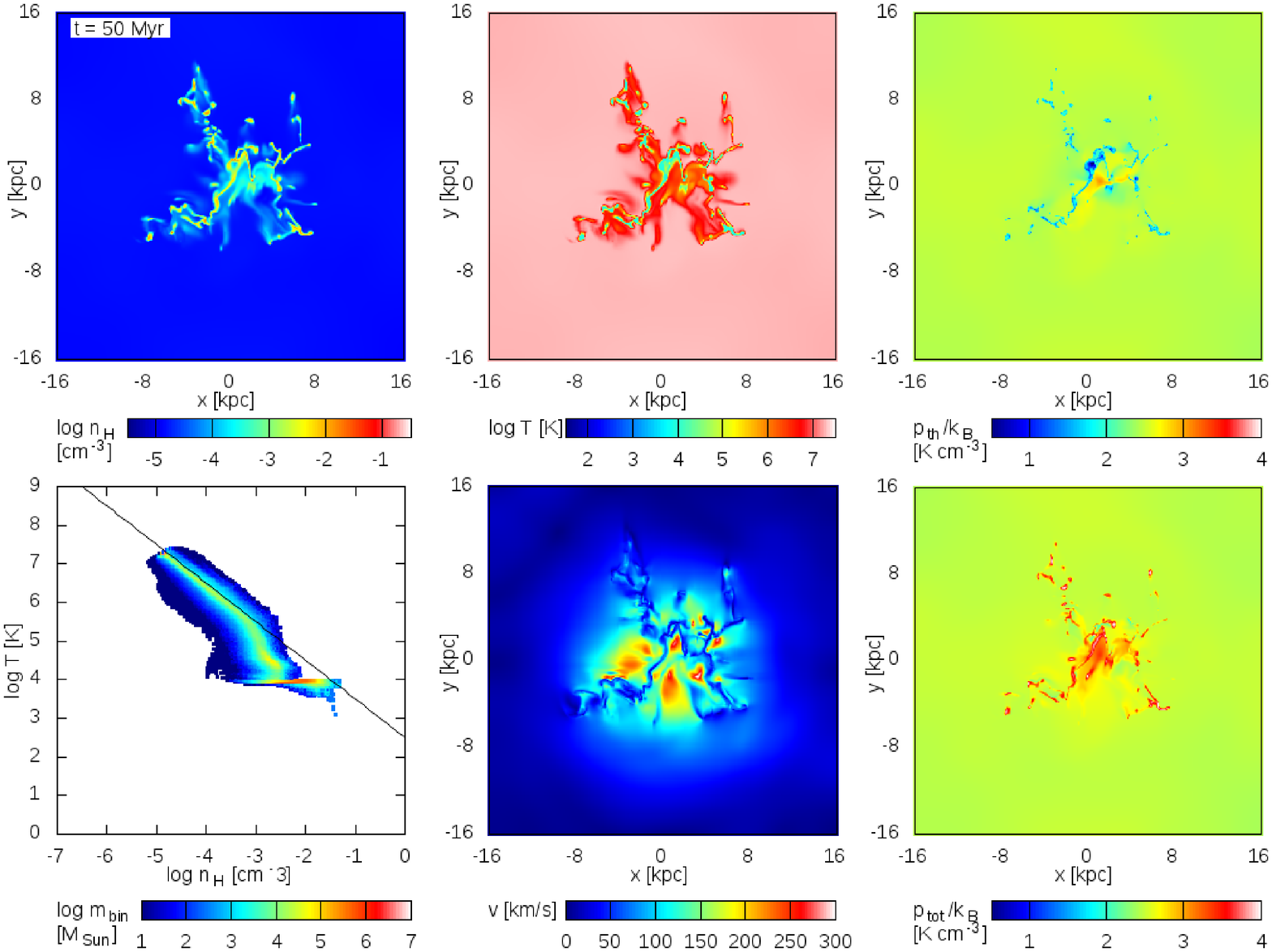}
\caption[hienv]{Snapshot of the major properties of the fiducial simulation at 50 Myr. Clockwise from top left : volume density slice; temperature; thermal pressure; total pressure; velocity; temperature-density plot.}
\label{fig:fidpara}
\end{figure*}

The evolution of the cloud is not completely dominated by its dynamics. A snapshot of the cloud's thermodynamic properties can be seen in figure \ref{fig:fidpara}. Hot gas stays close to the line of constant pressure (solid black in the bottom left panel). The majority of the sphere mass stays at around $\sim 10^4$\,K, because the cooling rate drastically drops at lower temperatures.

To ensure our results are not dependent on the resolution of the model, we re-ran the fiducial model but using a maximum resolution equivalent to 1024$^{3}$ grid cells instead of the standard 512$^{3}$. This run, model E, is otherwise identical to model A. The two models are extremely similar. Virtually the only difference is the S/N in the central beam : in both models this reaches a peak of about 8, but in model E it remains slightly higher throughout the rest of the simulation, with a final value of about 5 by the end of the simulation (compared to 4 for model A).The higher resolution allows the existence of small warm structures that cannot exist in the lower resolution simulations. Since this difference is very minor, and owing to the added computational costs of the higher resolution, we therefore used the standard resolution in all other simulations.

\begin{figure*}
\centering
\includegraphics[width=180mm]{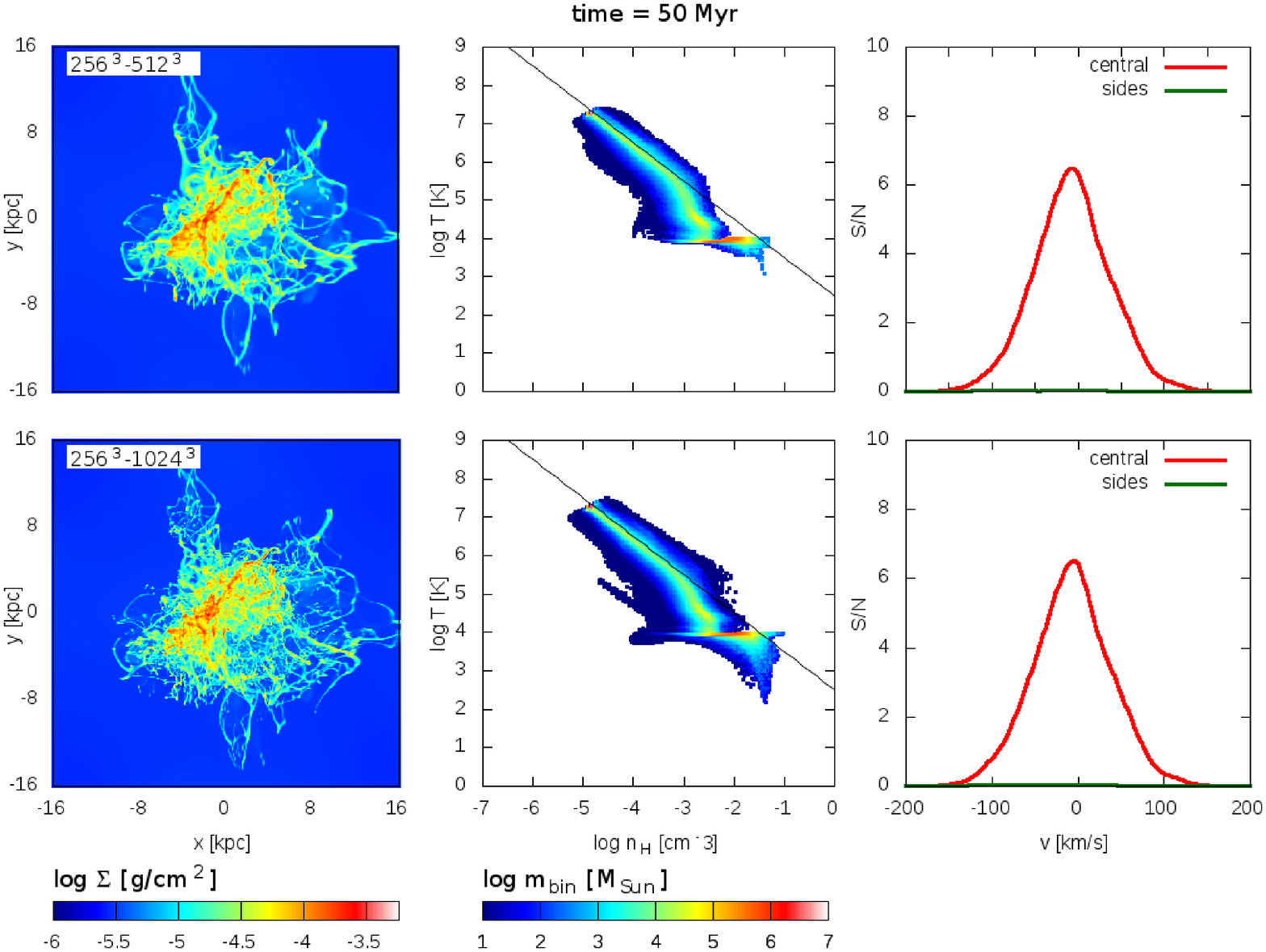}
\caption[hienv]{Comparison of the fiducial run (top row) with the high resolution run (bottom row).}
\label{fig:rescomp}
\end{figure*}

\subsection{Changing global cloud parameters}
\begin{figure}
\centering
\includegraphics[width=85mm]{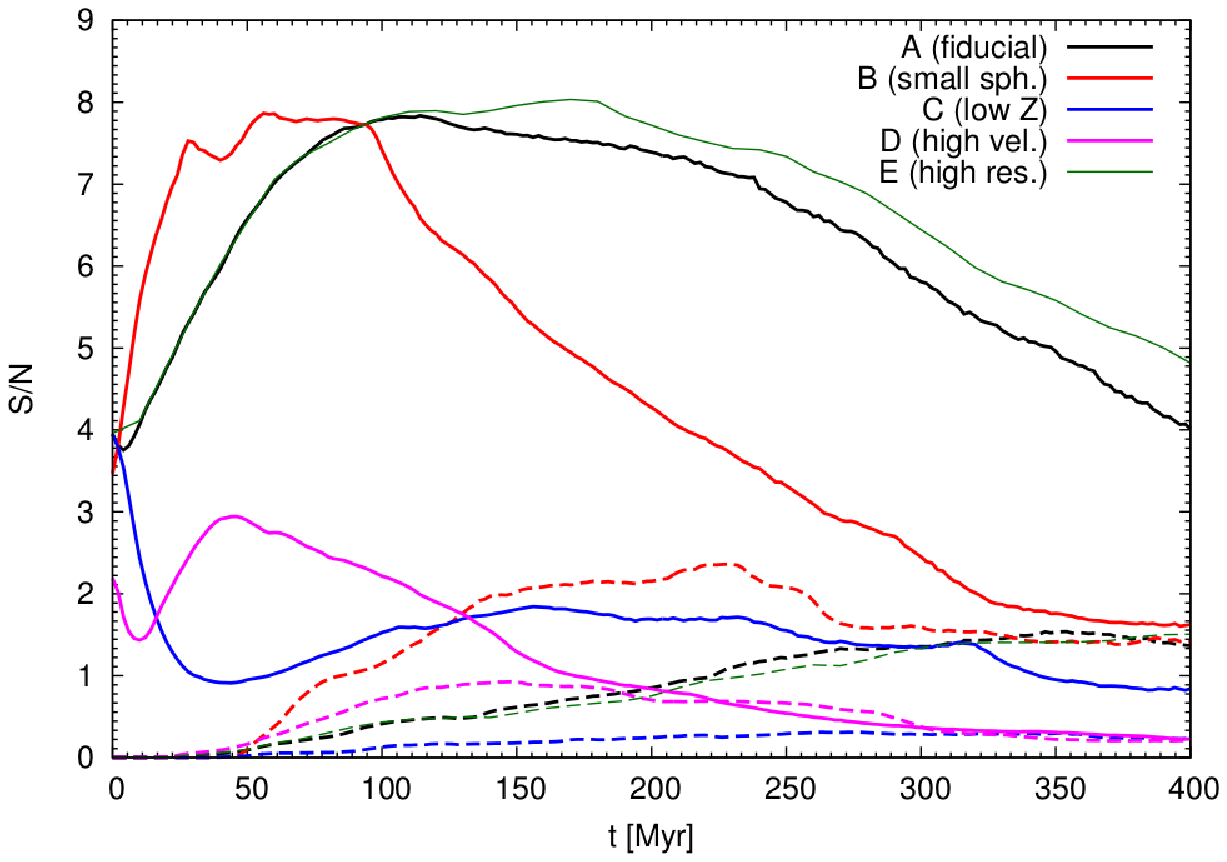}
\caption[hienv]{Comparison of models B, C, D and E S/N evolution with the fiducial model A. Solid lines indicate the central beam while dashed lines show the side beams.}
\label{fig:cloudssn}
\end{figure}

\begin{figure}
\centering
\includegraphics[width=85mm]{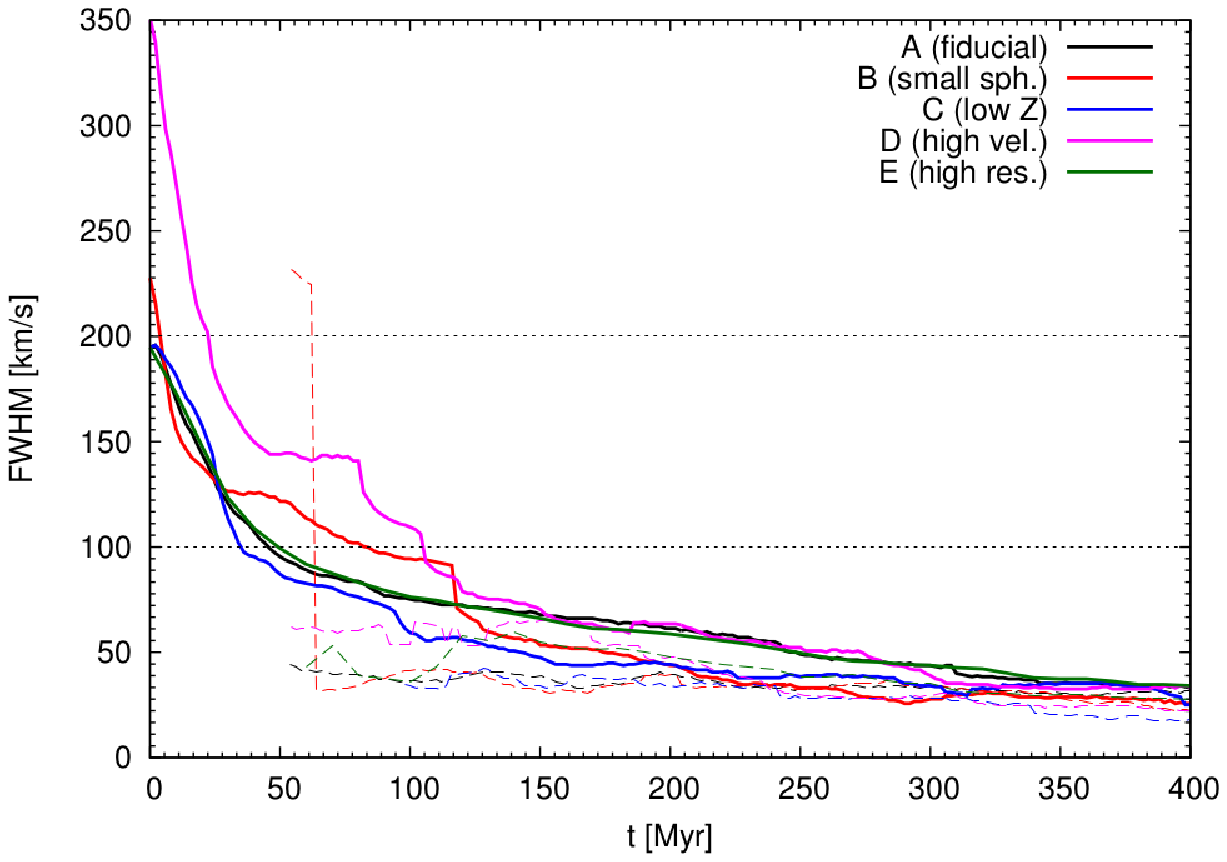}
\caption[hienv]{Comparison of models B, C, D and E FWHM evolution with the fiducial model A. Solid lines indicate the central beam while dashed lines show the side beams.}
\label{fig:cloudsfwhm}
\end{figure}

Models B, C and D each altered a single global parameter to investigate the effects independently. Based on those results, models J and K used additional adjustments to try and increase the time the clouds resembled the observations. The S/N and FHWM evolution can be seen in figures \ref{fig:cloudssn} and \ref{fig:cloudsfwhm}.

\subsubsection{The size of the cloud}
Reducing the size of the cloud to the smallest probable value of 2 kpc (model B) but retaining the FWHM does not prevent the cloud from dispersing, but does restrict the bulk of the mass to the confines of the central beam for a longer period. Internal collisions within the cloud act to slow the expansion and reduce the line width. As with the other dispersal cases, S/N increases, but not very dramatically, reaching a peak of about 8 as in the fiducial case. This happens more rapidly than in the fiducial run as the smaller crossing time of the cloud leads to the turbulence decaying more rapidly. The increase in S/N again comes at the expense of FWHM, which decreases rapidly. Since it has a slightly higher initial FWHM than the fiducial case, the FWHM does not fall below 100 \kms{} until 80 Myr. Although no longer matching the observations, the cloud remains detectable for approximately 200 Myr. 

\subsubsection{Cloud metallicity}
\begin{figure}
\centering
\includegraphics[width=85mm]{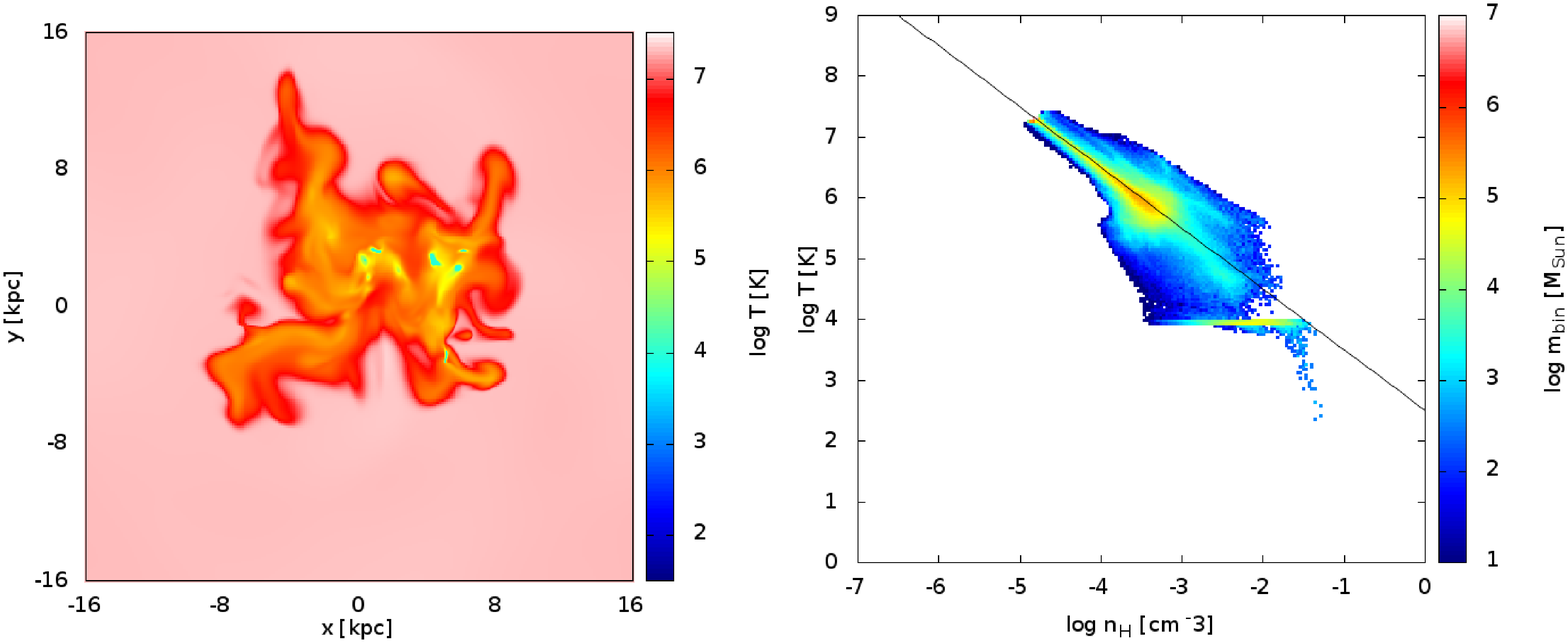}
\caption[hienv]{Temperature distribution within the central slice of the low metallicity (model C) cloud (left panel) and temperature-density distribution (right panel).}
\label{fig:modelc}
\end{figure}

Reducing the cloud metallicity to a primordial value (model C) changes the result entirely. Since the gas no longer cools efficiently, heating (mainly compressional, due to shocks created by the turbulence) increases the temperature of the gas extremely quickly and it effectively dissolves into the ICM. The S/N plummets, falling below 3 in 10 Myr. The FWHM also falls quickly, to a level below 100 \kms{} in 40 Myr. Snapshots of the temperature and density throughout this model are shown in figure \ref{fig:modelc}.

Interestingly, this heating is not seen when we use a smaller cloud with low metallicity (model K), at least not to the same degree. The higher density of the cloud allows it to still cool efficiently despite the low metallicity, radiating away most of its kinetic energy. Consequently the S/N still initially rises and drops, but the cloud remains detectable for around 130 Myr.

\subsubsection{Line width}
In model D we kept the turbulence parameters as in the fiducial case, but increased the ratio of kinetic to potential energy $Q$ to 3500 so as to increase the FWHM (approximately doubled). Unsurprisingly, this makes the cloud undetectable and it never exceeds a S/N of 3 during the 400 Myr of the simulation. However, it does have the benefit of increasing the FWHM and maintaining a high FWHM for a longer period.

Initially thinking that the cloud remained undetectable simply because there was less mass in each velocity channel, we tried doubling the mass of the warm gas (model J). This did not help. Although the S/N is initially a bit higher, it makes little difference to the simulation behaviour overall. It seems that the clouds strong internal motions cause significant heating, rendering it undetectable.

\subsection{Adjusting the turbulence}
\begin{figure*}
\centering
\includegraphics[width=180mm]{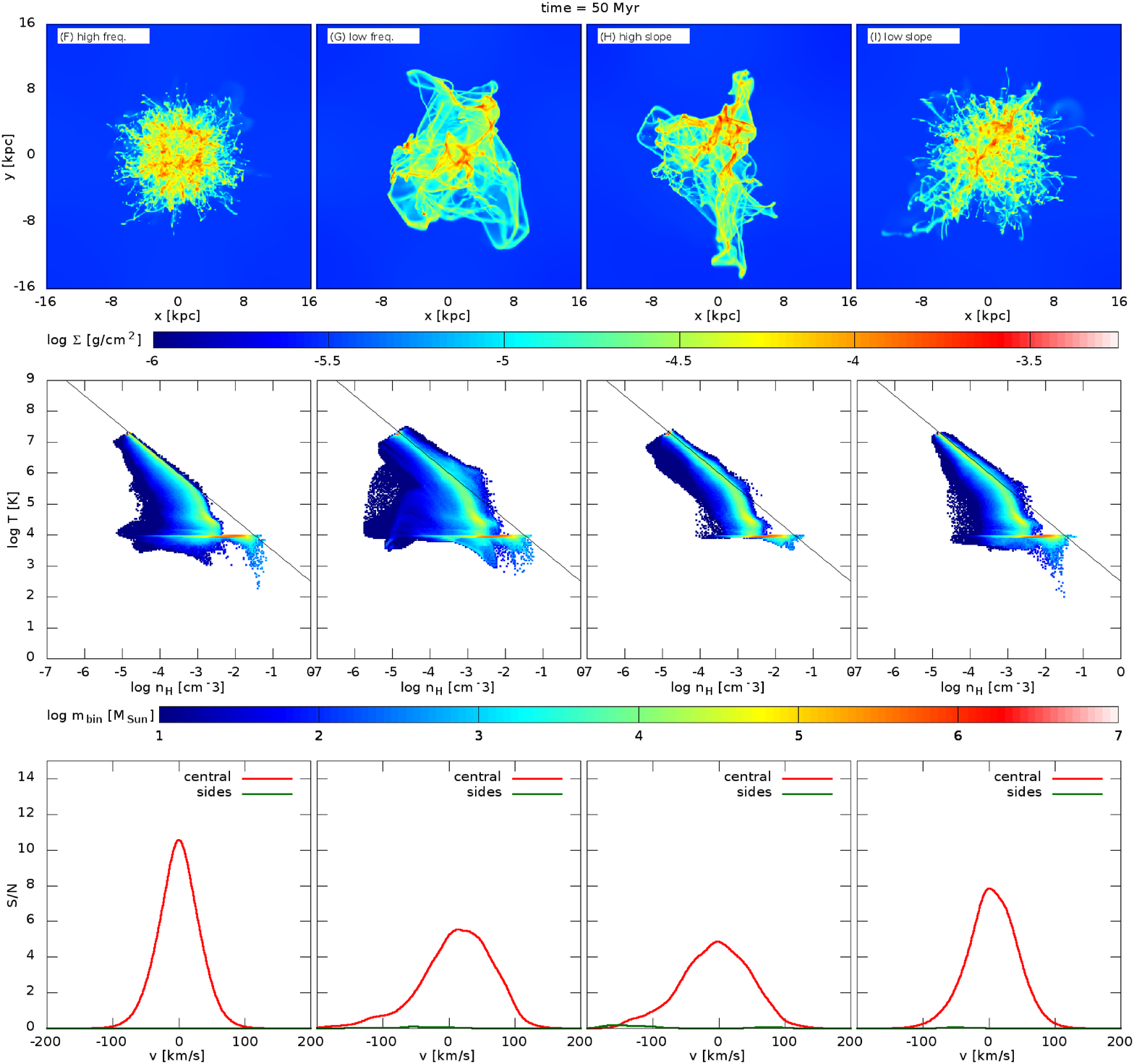}
\caption[hienv]{Comparison of the morphology and line profiles of models F-I, showing how altering the properties of the turbulence affects the behaviour of the cloud.}
\label{fig:turbcmp}
\end{figure*}

\begin{figure}
\centering
\includegraphics[width=85mm]{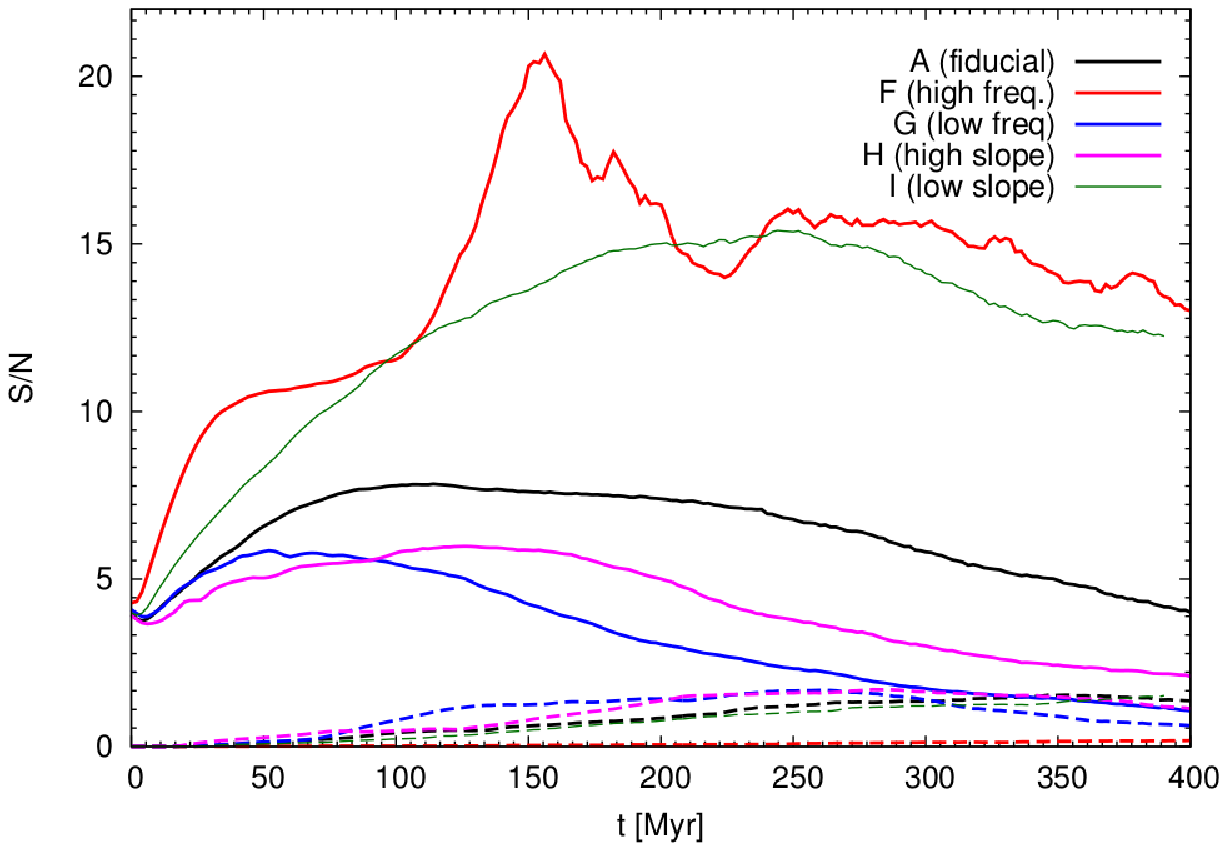}
\caption[hienv]{Comparison of models F-I S/N evolution with the fiducial model A. Solid lines indicate the central beam while dashed lines show the side beams.}
\label{fig:turbsn}
\end{figure}

\begin{figure}
\centering
\includegraphics[width=85mm]{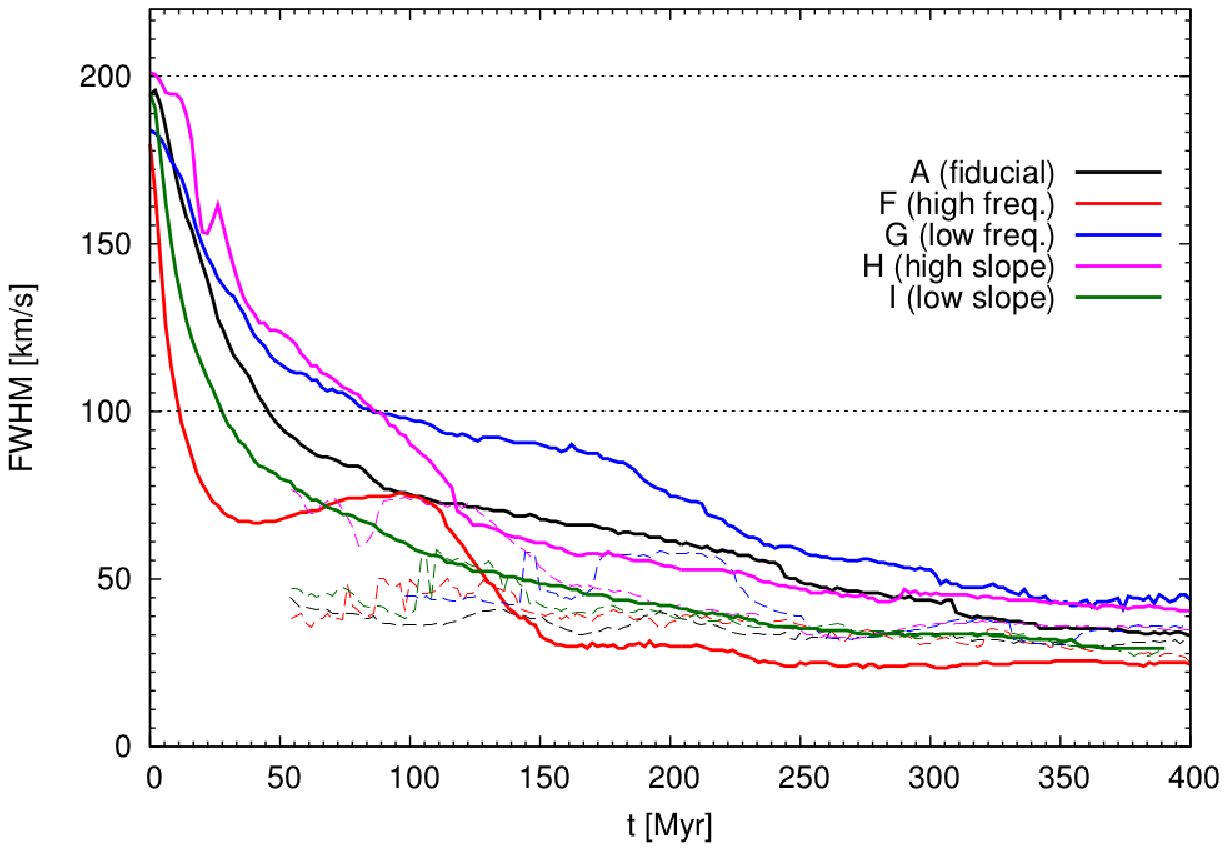}
\caption[hienv]{Comparison of models F-I S/N evolution with the fiducial model A. Solid lines indicate the central beam while dashed lines show the side beams.}
\label{fig:turbfwhm}
\end{figure}

Models F-I use identical parameters to the fiducial model, including the $Q$ ratio and FWHM, except for the specific properties of the velocity field. Interestingly, altering the slope and/or scale of the velocity variation can radically change the behaviour of the simulation, as illustrated in figure \ref{fig:turbcmp}. Unfortunately none of these alterations help the cloud more resemble the observed dark \HI{} clouds, but for different reasons. The S/N and FWHM evolution are shown in figures \ref{fig:turbsn} and \ref{fig:turbfwhm}.

As described we do not here include any source of energy to drive the turbulence, since we do not have a physical motivation for this. The crossing time for the 16 kpc and 4 kpc clouds is about 80 and 20 Myr respectively for the initial 200\,\kms{} respectively. These rather rapid crossing times mean the initial velocity fields will tend to dominate over any additional energy input, but again we emphasise that since the structures are gravitationally unbound, driven turbulence should only increase the dispersal rate. A caveat is that magnetic fields have been shown to be capable of providing extra stability in certain circumstances (\citealt{magnets}), but it is difficult to see how they could sustain the equilibrium state proposed in BL16. 

\subsubsection{Changing the spatial scale of the turbulence}
Increasing the spatial scale on which the velocity varies (model G) causes only minor morphological differences - a structure of long, thin, fragmenting filaments still forms, and they still disperse, except now the network is not apparent until about 30 Myr into the simulation. The S/N level evolves in a similar fashion to that of the fiducial model (see figure \ref{fig:turbsn}) though it is significantly lower than model A. This is perhaps because with the larger scale of the velocity variation, fewer substructures within the cloud collide with each other so the expansion is less hindered by internal motions. Additionally, the small dynamic range of the turbulence means it is not expected to provide any bulk support for the cloud. 

In contrast, reducing the scale of the velocity field (model F) causes very different behaviour. The long network of filaments is not seen, with the cloud instead transforming into smaller clouds which merge in the centre. Despite $Q$ being the same, and a few parts of the outskirts of the cloud escaping the simulation domain, the overall picture is one of collapse. As the cloud disintegrates into many small structures moving at random, it effectively collides with itself - preventing expansion. Cooling appears to be more significant, while the line width narrows dramatically (down to about 25 \kms{} by the end of the simulation) and S/N peaks at over 20. The observable properties of this model cease to resemble the real clouds in as little as 20 Myr, thanks to the rapid drop in FWHM.

\subsubsection{Changing the slope of the turbulence}
Changing the steepness of the gradient of the velocity variation has effects analogous to changing the scale. Model H, which has a steeper gradient than the fiducial run, has a similar outcome to model G (where the spatial scale length was increased). Both effectively reduce the number of regions within the cloud moving at different velocities, so the expansion proceeds unhindered by internal collisions. As shown in figures \ref{fig:turbsn} and \ref{fig:turbfwhm}, the evolution of models H and G is quantitatively and qualitatively similar.

The same is true for models F and I. Model F collapsed because the frequency scale of the turbulence meant there was a larger variation in the velocity from point to point within the cloud. Model I collapses because the shallower gradient of the turbulence causes much the same effect. The collapse of model F is rather stronger than I, which gradually begins to disperse by the end of the simulation, but both show a rapid drop in line width. Model I drops slightly more slowly than model F, but the line width is still below 100 \kms{} within 40 Myr. Neither resembles the observed clouds for very long.

\section{Summary and Discussion}
\label{sec:conc}
We performed the first numerical simulations of turbulent clouds as a proposed explanation for the optically dark AGES \HI{} sources. Such clouds have proved difficult to explain owing to their isolation ($>$\,100 kpc from the nearest galaxy), small size ($<$\,17 kpc diameter) and in particular their unusually high velocity widths ($\sim$\,150\,\kms{}). The explanation proposed by BL16 has the clouds being in pressure equilibrium with the ICM, and we made an attempt to numerically model this scenario. In our simulations, the line width of the clouds arises from dynamic turbulence, since the thermal pressure needed to generate line widths $>$\,100\,\kms{} requires higher temperatures than \HI{} can sustain. Our models show that this scenario is compatible with the lack of star formation (since they all retain low surface densities) and size of the clouds (none of our models would be detectable on scales $>$\,17 kpc), but cannot explain their other key features. We found that the high level of turbulence would mean the clouds are highly unstable, transient features, typically becoming undetectable on timescales $<$\,50-100 Myr. If this model is correct, it means that we have observed the six real clouds\footnote{These have only been detected in the 10\% of the cluster observed by AGES, suggesting that many more may be present elsewhere. We have recently begun an extension to AGES to search for clouds in a larger area of the cluster.} in a very short, unusual phase of their evolution. This would require an efficient formation mechanism, which we demonstrated in T17 is unlikely to be due to the effects of harassment. However, we also found that features of lower velocity width, more similar to the general population of known clouds, remained compatible with observations for much longer periods.

Our models showed three basic modes of behaviour of the clouds, depending on their precise initial conditions : dispersal, collapse, and rapid heating. All models showed a common evolution from a homogeneous-density sphere into a network of fragmenting filaments. Regardless of the metallicity, cloud mass, line width or parameters of the turbulence, the clouds never exist in any sort of quasi-equilibrium state : they are inherently and extremely unstable. The closest we came to achieving any kind of stability was using a strong variation of the internal motions within the cloud, in essence transforming it into a series of smaller clouds moving in random directions. Collisions of these clouds causes a loss of kinetic energy, so that during its early evolution the structure overall collapses (despite being initially strongly gravitationally unbound considering only the ratio of kinetic to gravitational potential energy). However even these cases start to disperse by the end of the simulation, and very rapidly lose the high line width that originally identified the observed clouds as interesting features.

The model which best reproduced the observed parameters for the longest possible time was model B, which is compatible with the observations for about 80 Myr. It is possible that with further adjustments we might find a solution that gives agreement for a longer time, but this then requires fine-tuning without physical justification : there is no known reason to suspect that stripped gas should naturally form 2\,kpc spherical clouds with a very narrow range of turbulent velocity parameters.

In T16 we made a simple analytic estimate of the cloud's lifetimes (meaning how long they remain compatible with the observations of both S/N and line width) assuming their measured velocity width indicates their expansion velocity. Neglecting self gravity and the ICM, this gave a value of 125 Myr assuming the clouds to have initial radii of approximately 2 kpc. After this the clouds would be detectable but larger than the Arecibo beam, so in disagreement with the observations. Here our much more sophisticated numerical models, which account for heating and cooling, metallicity, turbulence, self-gravity and pressure confinement, show that this analytic value is likely an overestimate. This more realistic model shows that the ICM is likely to reduce the lifetime of the high width clouds, not increase it.

The reason this approach does not seem tenable is because the turbulent motions needed to provide pressure support are not equivalent to thermal pressure. Instead of providing a uniform support against the cloud's self-gravity and the pressure from the ICM, which for thermal pressure \cite{bell} have demonstrated is possible for low velocity width clouds, turbulence creates inherently unstable, changing structures which quickly destroy the unfortunate clouds. Specifically, the initial sphere has an unstable surface. Any parcel of gas with a slightly higher than average density and consequently higher ram pressure can expand into the ambient medium. Conversely, regions with lower density and the ram pressure around it are compressed by the ambient hot gas, which penetrates into the volume of the sphere. As a result, the original overdensity soon becomes a separate cloud of a gas embedded in the almost uniform hot ICM. In this way, the turbulent sphere quickly decomposes into a set of smaller clouds, with their internal pressure being in equilibrium with the thermal pressure of the ICM. The clouds move almost freely, because the pressure force from the ICM is almost the same from all directions, and because the gravitational force is negligible.

In T16 and T17 we considered how the clouds might form due to tidal encounters between galaxies. We found that this was possible for the clouds with low line widths ($\sim$\,30\,\kms{}) but not for those with the high line widths. Here we did not explicitly investigate the low-width clouds, but, interestingly, found that they formed quite readily in many models and are apparently long-lived. Similarly, we also showed in T16 that such features are compatible with the interpretation of the clouds being rotating dark galaxies; in that scenario the low-width clouds would be rotating discs seen close to face-on. There is essentially no problem in explaining clouds of low line width, at least in this environment~: rather the difficulty would be in establishing which (if any) of these possibilities is correct. Together with the models of \cite{bell}, these simulations support that pressure confinement may be extremely significant for the long-term survival of low width clouds, which we emphasise are much more common than high width objects (see T16 for a review). These numerical results can be used to help distinguish between the effects of external pressure and tidal encounters, potentially giving a new interpretation to many previously documented features.

BL16 calculate that the low velocity width clouds described in \cite{cannon} and \cite{jan} are, if in apparent pressure equilibrium with the surrounding medium, close to the limit where their density would be sufficiently high for H$_{2}$ formation. Furthermore, the \cite{bell} cloud has a faint stellar counterpart with evidence for recent star formation. However, BL16 note that the AGES clouds would have a density well below the H$_{2}$ formation threshold, and though our models here do sometimes show localised density increases, none of the gas approaches this level (bearing in mind the caveats that we do not have an accurate chemical network and do not account for heat conduction).

The initial collapse, and reduction of velocity width, of some of our models raises the intriguing possibility that the high width clouds could be the progentiors of the low width clouds. Unfortunately, this idea has some fundamental difficulties - at least as regard to the specific AGES clouds as a population. The six high velocity width AGES clouds outnumber the two low width AGES clouds in the same area, whereas the dynamics would suggest that the low width phase should be both longer lived and easier to detect due to the higher S/N ratio. This problem could be overcome if the phase change from \HI{} to H$_{2}$ is short (compared to the time for their initially high velocity width to fall), thus quickly rendering the low velocity width clouds undetectable and perhaps initiating star formation. A further difficulty is that our simulations suggest that the high width phase is itself very brief (see below) and therefore it is improbable that we detect any clouds at all in the even shorter low width phase of their evolution. More generally, regardless of the line widths of the clouds under investigation, these results also demonstrate the importance of considering the effects of the ICM in their evolution.

Our attempts in T16, T17 and the present work show that both harassment and the ICM cause important, different effects on the evolution of clouds in the absence of each other. In reality neither proceeds with the other so entirely suppressed, so future models should examine the combination of the two. Our main conclusion thus far is that low width clouds are readily explicable by different, separate mechanisms, while high width clouds require a more complex formation process. This could either be through a combination of physical processes (not only the ICM pressure and tidal harassment, but also rotation, magnetic fields, and thermal effects from the ICM), and/or (more radically) because they have a fundamentally different nature.

We considered the possibility in T16 that the high width clouds could be dark galaxies in which their line width arises due to rotation. We showed that such objects would be stable to the effects of harassment in the Virgo cluster and would retain their measured FWHM and S/N even on 5 Gyr timescales. This is a tempting interpretation given recent discoveries of large numbers of very low surface brightness galaxies, however there are two important caveats. First, our results here demonstrate the importance of the ICM on low-mass objects, and for discs moving through the cluster the effect of ram-pressure stripping could be even more significant (we will examine the surviability of dark galaxies in the ICM in future simulations). Second, we have recently obtained VLA data of all of the clouds, and with this improvement in resolution, we expect to soon be able to distinguish between all the suggested interpretations. Our models described here allow us to predict in advance exactly what we should detect (pending data reduction) if we really are witnessing multiple clouds in a brief, peculiar phase of their evolution, either due to harassment or from collapsing turbulent structures. While it seems (in our view) unlikely that we would be happening to witness several clouds during a short phase of their evolution, the only way to really distinguish between the various scenarios is through higher resolution observations.

\section*{Acknowledgments}

We thank the anonymous referee for their helpful comments and suggestions which improved the manuscript. We also wish to thank Blakesley Burkhart, Abraham Loeb and Lukas Leisman for helpful discussions regarding the nature of the AGES \HI{} clouds and possible deviations from the Tully-Fisher relation.

This work was supported by the Albert Einstein Center for Gravitation and Astrophysics via the Czech Science Foundation project 14-37086G, the institutional project RVO 67985815, and the Czech Ministry for Education, Youth and Sports research infrastructure grant LM 2015067. We also thank King Henry II of England for unwittingly providing the title via Robert Minchin.

This work is based on observations collected at Arecibo Observatory. The Arecibo Observatory is operated by SRI International under a cooperative agreement with the National Science Foundation (AST-1100968), and in alliance with Ana G. M\'{e}ndez-Universidad Metropolitana, and the Universities Space Research Association.

{}

\end{document}